\documentclass{jaa}
\usepackage{natbib}

\usepackage{graphicx}
\usepackage{float}
\usepackage{booktabs}
\usepackage{longtable}
\usepackage{placeins}
\usepackage{array}
\usepackage{enumitem}
\usepackage{comment}
\usepackage{hyperref}
\usepackage{booktabs}
\usepackage{amssymb}

\begin{document}\sloppy

\title{Searching for a superdisk in radio galaxy J0116-473}


\author{Ankur Sinha\textsuperscript{1*}, Riya Rathore\textsuperscript{1}, Narendra Nath Patra\textsuperscript{1} and Abhirup Datta\textsuperscript{1}}
\affilOne{\textsuperscript{1}Department of Astronomy, Astrophysics and Space Engineering, Indian Institute of Technology Indore, 453552, India.\\}


\twocolumn[{

\maketitle

\corres{ms2404121004@alum.iiti.ac.in}


\begin{abstract}
Superdisks have emerged as an active area of research in recent years, and J0116$-$473 represents a promising target for studying this extended structure. Our primary objective was to search for H\,{\sc i} absorption associated with the suspected superdisk. However, no such absorption feature was detected, suggesting a low, or absence of neutral hydrogen content in the superdisk. In addition, we examined a compact point source located near the galaxy's core and the presumed plane of the superdisk, enabling us to search for H\,{\sc i} absorption against this background continuum. We also present a detailed multi-band morphological analysis of the galaxy using Giant Metrewave Radio Telescope (GMRT) observations in Bands~3, 4, and~5. A spectral analysis of both the galaxy and the nearby point source was carried out using data from these three frequency bands. A systematic steepening of the spectral index is observed from the core toward the lobes, as expected for aging synchrotron-emitting plasma. We also found that the northern inner lobe exhibits a significantly steeper spectrum than its southern counterpart, possibly reflecting environmental effects associated with the proposed superdisk. Since superdisks are expected to contain hot, ionized gas, we additionally examined archival X-ray observations from the \textit{XMM--Newton} telescope. Although diffuse X-ray emission associated with the radio lobes is visible, no significant emission is detected from the region corresponding to the suspected superdisk.  
\end{abstract}


}]


\doinum{12.3456/s78910-011-012-3}
\artcitid{\#\#\#\#}
\volnum{000}
\year{0000}
\pgrange{1--}
\setcounter{page}{1}
\lp{1}

\section{Introduction}

Superdisks are large disc-like structures surrounding the elliptical host galaxy and oriented nearly perpendicular to the radio jets. They are thought to consist of large-scale gaseous structures associated with the host galaxy, possibly containing thermal gas and dust \citep{GopalKrishna2004, GopalKrishna2008}. At radio wavelengths, they appear as strip-like emission gaps between the lobes. The key observational signature of a superdisk is a broad, elongated emission gap with sharp, nearly linear inner edges located between the collimated radio lobes of powerful radio sources \citep{GopalKrishna2008}. These structures, described as ‘broad pancakes’ by \citep{GopalKrishna2000}, typically have diameters of $\geq 75\, \mathrm{kpc}$ and widths of $\approx 25\,\mathrm{kpc}$. Several formation scenarios have been proposed for these putative superdisks, including the tidal capture of gas-rich galaxies and the remnants of damped Lyman-$\alpha$ clouds; however, direct observational evidence for their origin remains unclear \citep{GopalKrishna2000}.\\

The composition of superdisks has been investigated using several observational approaches, although it remains poorly constrained. X-ray observations of some superdisk candidates indicate the presence of hot ionized gas \citep{Croston2005}, while other studies suggest that a cooler, dusty component may also be associated with these structures\citep{deKoff2000, Temi2004}. Sensitive H\,{\sc i} 21-cm observations have been further used to constrain the neutral hydrogen content of superdisks. In particular, a comprehensive search for H\,{\sc i} 21-cm emission from superdisks in four radio galaxies (3C227, 3C98, 4C+32.25, and 3C192) done by \citep{Anand2019}, yielded no detections, placing stringent upper limits on their H\,{\sc i} content. These results rule out an H\,{\sc i}-dominated superdisk as a viable model; however, the possibility that superdisks contain low-density neutral hydrogen below the current detection limits, or are composed primarily of warm or hot gas, remains open \citep{Anand2019}.\\

To explain the formation of superdisks, several scenarios involving different physical processes have been proposed:

\begin{enumerate}

\item \textbf{Merger-driven formation:}
One widely discussed scenario involves a merger between the giant host galaxy and a smaller elliptical galaxy. In this picture, gas associated with the infalling galaxy transfers its angular momentum to the host galaxy's circumgalactic medium, causing it to flatten into a large, pancake-like structure, identified as the superdisk \citep{GopalKrishna2008}. The gravitational recoil from the merger may also account for the host galaxy's asymmetric position relative to the superdisk. Extended hot gaseous halos around elliptical galaxies provide a plausible reservoir for this process \citep{Mathews2003}, while numerical simulations demonstrate that galaxy mergers can efficiently redistribute angular momentum within gaseous components \citep{Barnes1996}.

\item \textbf{Jet–wind interactions:}
Another proposed mechanism involves the dynamical interaction between strong thermal winds emerging from the active galactic nucleus (AGN) and the back-flowing synchrotron plasma from the radio lobes. This interaction becomes particularly relevant once the relativistic jets propagate beyond the wind zone and enter the surrounding intergalactic medium \citep{GopalKrishna2007}. Backflows of relativistic plasma in radio lobes are well established from both observations and simulations \citep{Norman1982, Leahy1984}, while AGN are also known to drive powerful thermal winds capable of influencing their surrounding environments \citep{King2015}.

\item \textbf{Spin-flip and black hole merger events:}
A third scenario invokes rapid precession of the supermassive black hole following a major merger. In this framework, the final stages of a black hole merger can lead to a spin-flip of the black hole axis, resulting in rapidly precessing jets \citep{Merritt2002, GopalKrishna2003}. These swinging jets may entrain surrounding gas and form a narrowing, cone-like outflow, which could eventually produce a superdisk-like structure \citep{Gopal-Krishna2012}. Precessing jets are also a natural consequence of binary black hole systems in active galactic nuclei \citep{Begelman1980}, and jet–environment interactions can further influence the surrounding gaseous medium through entrainment processes \citep{Bicknell1994}.

\end{enumerate}

The superdisk model has been invoked to explain several puzzling astrophysical phenomena, including the lobe depolarization asymmetry of quasars, the correlated radio–optical asymmetry of radio galaxies (hereafter RGs), low-frequency variability of high-latitude blazars, and absorption features in the Ly$\alpha$ profiles of small RGs \citep{GopalKrishna2000}.\\
Several fundamental asymmetries observed in powerful radio galaxies have been interpreted within the superdisk model. In particular, differential depolarization produced by magnetized hot plasma covering one radio lobe more than the other has been proposed to explain the Laing–Garrington effect, which manifests as an asymmetry in the polarization fraction between the jet and counterjet sides \citep{Garrington1988}. An asymmetrically positioned superdisk can naturally produce unequal depolarization across the two lobes, thereby accounting for this effect \citep{Garrington1991}.\\
Superdisks have also been invoked to explain the remarkably correlated radio–optical asymmetries observed in FR II galaxies, where the brighter [O II] $\lambda3727$ emission is found on the same side as the brighter radio hotspot \citep{McCarthy1991,McCarthy1993}. In this picture, asymmetric energy deposition preferentially ionizes gas on one side of the host galaxy, producing the observed optical emission. At the same time, the jet propagating through a smaller column of superdisk material advances farther and forms a brighter hotspot, as the superdisk forces the two jets to traverse unequal path lengths through the surrounding dense medium \citep{GopalKrishna2004, Bell2013, GopalKrishna2008}.\\

Superdisks can significantly influence the dynamics of jet propagation. Jets propagating through superdisk material traverse different path lengths, leading to varying degrees of entrainment and differential deceleration \citep{GopalKrishna2008}. The jet that encounters a larger column of superdisk material undergoes stronger entrainment and decelerates more rapidly, whereas the opposing jet, interacting with less material, experiences weaker confinement and retains a higher velocity \citep{Bicknell1994}. This asymmetric propagation directly affects the locations of the hotspots and the efficiency of particle acceleration.
Superdisk–jet interactions can also influence how energy is deposited into the surrounding environment. When jets interact with superdisk material, enhanced shock formation can lead to more efficient particle acceleration and greater energy dissipation into the ambient medium \citep{GopalKrishna2000,GopalKrishna2007,Begelman1989}. Furthermore, by guiding the propagation of jets within their plane, superdisks may facilitate the transport of large-scale magnetic fields into the intergalactic medium, possibly through the circumgalactic medium \citep{Perucho2014, GopalKrishna2007, GopalKrishna2008}.\\

X-shaped radio galaxies (XRGs), characterized by two misaligned pairs of radio lobes that form a distinctive X-shaped morphology, may be physically connected to the formation of superdisks \citep{GopalKrishna2000,GopalKrishna2003}. One proposed explanation links both phenomena to mergers involving supermassive black holes in galactic nuclei. Supporting evidence comes from shell structures observed in several XRG host galaxies, which are widely interpreted as fossil remnants of past galaxy mergers produced by oscillating stellar populations following the merger event \citep{Malin1983,Hernquist1988}. In this scenario, the coalescence of misaligned supermassive black holes leads to a rapid reorientation of the remnant black hole's spin axis \citep{Merritt2002}. The resulting spin flip causes the jets to change direction and interact with the pre-existing superdisk material, potentially producing X- or Z-shaped radio structures \citep{GopalKrishna2000, GopalKrishna2003, Merritt2002, DennettThorpe2002, capetti2002origin, Begelman1980}.
Recent simulations, however, suggest that jet–gas interactions alone may generate X-shaped morphologies without requiring black hole mergers \citep{capetti2002origin}. In particular, powerful jets can be deflected into X-shaped configurations when they propagate through anisotropic pressure environments associated with superdisks or triaxial gaseous distributions \citep{Kraft2005, HodgesKluck2011, Rossi2017}. These results indicate that superdisks may play an important role in shaping a variety of radio galaxy morphologies by influencing both jet propagation and the surrounding gaseous environment \citep{Giri2024, GopalKrishna2008}.\\

The radio galaxy J0116-473, also known as PKS 01147-47, lies at a redshift of $z = 0.146$ and has a projected linear size of $\approx1.8\,\mathrm{Mpc}$ \citep{Dabziger1978}. Its celestial coordinates are RA $= 01^{\rm h}16^{\rm m}25.04^{\rm s}$ and Dec $= -47^\circ22'41.6''$ (J2000). This source is classified as a double-double radio galaxy (DDRG) \citep{Schoenmakers2000}, characterized by two distinct pairs of radio lobes on opposite sides of the host galaxy. In general, radio galaxies exhibiting a single pair of lobes are referred to as single–double radio galaxies (SDRGs), while systems containing three pairs of lobes are known as triple–double radio galaxies (TDRGs) \citep{Brocksopp2007, Hota2011}.
Each pair of lobes corresponds to a separate episode of jet activity from the active galactic nucleus (AGN) \citep{Konar2013}. Consequently, the presence of two pairs of lobes in a DDRG indicates that the galaxy has undergone at least two distinct cycles of jet activity.
\citep{Konar2013} performed a spectral ageing analysis of J0116-473 and derived approximate constraints on the timescales associated with its episodic activity. They estimated the age of the inner double ($t_{\rm innd}$) to lie within $1 \lesssim t_{\rm innd} \lesssim 28\,\mathrm{Myr}$, while the age of the outer double ($t_{\rm outd}$) was found to be $66 \lesssim t_{\rm outd} \lesssim 236\,\mathrm{Myr}$. The age of the warm spot in the outer lobes ($t_{\rm ws}$) was constrained to $t_{\rm ws} \lesssim 64\,\mathrm{Myr}$, and the travel time of the jet from the nucleus to the hotspot ($t_{\rm jet}$) was estimated to be $\approx2.4\,\mathrm{Myr}$.
Radio galaxies exhibiting episodic jet activity generally undergo two main evolutionary phases: an active phase, during which relativistic jets are launched from the nucleus, and a quiescent phase, when jet production ceases. For J0116-473, \citep{Konar2013} estimated the duration of the quiescent phase to be $1.4 \lesssim t_{\rm quies} \lesssim 65.4\,\mathrm{Myr}$, while they constrained the duration of the active phase to $t_{\rm active} \lesssim 170\,\mathrm{Myr}$. \\

The radio galaxy J0116-473 provides an excellent opportunity to investigate the presence of a superdisk because of its favorable geometry and morphology. In this work, we attempt to detect the superdisk in this source through observations of its H\,{\sc i} absorption spectrum. For this purpose, we analyze data obtained with the GMRT in Bands 3, 4, and 5, corresponding to observing frequencies of 325, 610, and 1240 MHz, respectively. The L-band observation centered at 1240 MHz is used to search for H\,{\sc i} absorption against the point source located in the plane of the superdisk. These multi-frequency observations allow us to examine the source morphology across different radio bands.\\

\section{ Sample Selection}
\label{sec:sample_selection}

The radio galaxy J0116-473, also known as PKS 0114-47, is a particularly favorable target for this study due to its rare geometry and well-characterized radio structure. It is one of only two known systems in which a bright non-thermal radio component is observed through a superdisk, providing a unique opportunity to directly probe the H\,{\sc i} content of such structures. As a double–double radio galaxy (DDRG), the synchrotron emission from its inner lobes passes through the superdisk, allowing any detected H\,{\sc i} absorption to be unambiguously attributed to the superdisk rather than to gas associated with the host galaxy.
The source also provides a suitable background continuum for absorption studies, with a total flux density of $\approx 0.26$ Jy at 1.4 GHz, of which about 120 mJy originates from the lobe projected against the superdisk \citep{Saripalli2002}. Its clearly defined radio lobes and the central emission gap further support the presence of a superdisk. Moreover, at a redshift of $ z\approx 0.146$ (which corresponds to a luminosity distance of $\approx 650 \,\mathrm{Mpc}$), J0116-473 is too distant for practical H I emission studies. Still, it is well suited for absorption-line spectroscopy, as the redshifted 21-cm line falls within the L band accessible to current radio telescopes.

\begin{table*}[t]
\centering
\small  
\setlength{\tabcolsep}{4pt} 
\renewcommand{\arraystretch}{1.2}

\newcolumntype{L}[1]{>{\raggedright\arraybackslash}p{#1}}
\newcolumntype{C}[1]{>{\centering\arraybackslash}p{#1}}
\newcolumntype{R}[1]{>{\raggedleft\arraybackslash}p{#1}}

\begin{tabular}{
    L{1.8cm}  
    L{3.0cm}  
    L{4.0cm}  
    L{3.5cm}  
    L{2.2cm}  
}
\toprule
\textbf{Proposal Code} & \textbf{PI\_Name} & \textbf{Time on Source} & 
\textbf{Observation Central Frequency (Band)} & \textbf{Channel Width} \\
\midrule

13JMA01 & J. Machalski &
\begin{minipage}[t]{4.3cm}
\begin{itemize}[leftmargin=*, itemsep=0pt, topsep=1pt]
    \item 07 Mar 2008: 3 hr 45 min
    \item 14 Mar 2008: 3 hr 46 min
\end{itemize}
\end{minipage}
&
\begin{minipage}[t]{3.8cm}
\begin{itemize}[leftmargin=*, itemsep=0pt, topsep=1pt]
    \item 610 MHz (Band 4)
    \item 325 MHz (Band 3)
\end{itemize}
\end{minipage}
&
\begin{minipage}[t]{2.5cm}
\begin{itemize}[leftmargin=*, itemsep=0pt, topsep=1pt]
    \item 125 kHz
    \item 125 kHz
\end{itemize}
\end{minipage} \\
\midrule

32\_087 & Narendra Nath Patra &
\begin{minipage}[t]{4.3cm}
\begin{itemize}[leftmargin=*, itemsep=0pt, topsep=1pt]
    \item 18 Jun 2017: 1 hr
    \item 21 Jun 2017: 2 hr 20 min
    \item 23 Jun 2017: 2 hr
\end{itemize}
\end{minipage}
&
\begin{minipage}[t]{3.8cm}
\begin{itemize}[leftmargin=*, itemsep=0pt, topsep=1pt]
    \item 1250 MHz (Band 5)
    \item 1240 MHz (Band 5)
    \item 1240 MHz (Band 5)
\end{itemize}
\end{minipage}
&
\begin{minipage}[t]{2.5cm}
\begin{itemize}[leftmargin=*, itemsep=0pt, topsep=1pt]
    \item 400 kHz
    \item 200 kHz
    \item 200 kHz
\end{itemize}
\end{minipage} \\

\bottomrule
\end{tabular}
\caption{Observation details for J0116–473 obtained from GMRT archival data.}
\label{tab:gmrt_obs}
\end{table*}

\section{Observation and Data Analysis}
We used observations of J0116-473 obtained with the GMRT in three frequency bands: Bands 3, 4, and 5. The Band 5 observations were conducted at a central frequency of 1240 MHz with a total observing time of $\approx 8\,\mathrm{h}$, of which $\approx 5\,\mathrm{h}$ were spent on source, under project code $32\_087$. In addition, we used archival GMRT data from project code 13JMA01 for Bands 3 and 4, observed at 325 and 610 MHz, respectively.
For Band 5, observations were carried out on three different days, as listed in Table~\ref{tab:gmrt_obs}. However, we used only two of these datasets for imaging. Due to increased radio frequency interference, we excluded the data from 18th July. Moreover, including the 18 June data does not significantly improve the final image's sensitivity. Therefore, the Band 5 imaging was performed using only the data from 21 and 23 June. \\


The Band~4 data were recorded in two spectral windows. One window was severely affected by radio-frequency interference (RFI) that could not be removed through flagging; therefore, only the remaining spectral window was used for further analysis.
For Band 5 observations, we used data obtained with the GWB (GMRT Wideband Backend) \citep{Reddy2017}, while the Band 3 and Band 4 data were acquired using the GSB (GMRT Software Backend) \citep{roy2010}. Band 3 and 4 observations were done before the GWB was implemented at GMRT. The Band 5 observation was performed with a total bandwidth of $200\,\mathrm{MHz}$ and 4096 spectral channels, resulting in a spectral resolution of $\approx 390\,\mathrm{kHz}$ (corresponding to a velocity resolution of $\approx 92\,\mathrm{km\,s^{-1}}$).
The Band 3 and Band 4 observations were performed with a bandwidth of $32\,\mathrm{MHz}$ and 256 channels, resulting in a spectral resolution of $\approx 125\,\mathrm{kHz}$ (corresponding velocity resolution of $\approx 29\,\mathrm{km\,s^{-1}}$), Standard flux calibrators such as 3C48, 3C147, and 3C286 were used for flux and bandpass calibration during each observing run. A nearby phase calibrator located within $\lesssim 10^\circ$ of the target source was observed approximately once every 45 minutes for phase calibration.\\

All data reduction was carried out using the Common Astronomy Software Applications package (\texttt{CASA}; \citealt{CASA2022}), following standard flagging and calibration procedures. At the beginning of the analysis, dead antennas and defective baselines were identified and flagged. Phase calibrators were used to determine the phase solutions, while a primary calibrator was used for flux density calibration. Bandpass calibration was performed using the same primary calibrators.\\
After flagging and calibration were applied, a separate measurement set (MS) containing only the target source (J0116-473) was created by removing the calibrator data. The raw visibility data for Band 5 consisted of 4096 frequency channels. To improve computational efficiency and increase the signal-to-noise ratio per channel, the data were spectrally averaged by a factor of eight, resulting in 512 channels in the final dataset. The Band 3 and Band 4 datasets contain only 256 channels each and were therefore not spectrally averaged.\\
Spectral imaging was performed using the \texttt{tclean} task in \texttt{CASA}, while continuum imaging was carried out using \texttt{WSClean} \citep{offringa2014}. Full-resolution images were produced in all three frequency bands using Briggs robust weighting ($\texttt{ROBUST} =  0.5$) \citep{briggs1995}. The pixel scale was chosen such that each synthesized beam was sampled by at least three pixels.
Owing to poor phase calibration in the Band 3 data, the resulting images were not properly aligned with the Band 5 images. To correct this, we performed a manual astrometric alignment. Four compact point sources common to both images were selected, and the coordinates of their peak flux densities were determined using the \texttt{imfit} task of \texttt{CASA}. The positional offsets between the Band 3 and Band 5 images were then calculated from these measurements, and the derived offset was applied to the Band 3 image to achieve proper alignment.\\


Furthermore, we analysed archival \textit{XMM--Newton} observations of J0116$-$4722 (ObsID 0601260101), obtained on 28 November 2009, to investigate its X-ray morphology. The pointing was centred at RA $= 01^{\rm h}16^{\rm m}25.04^{\rm s}$ and Dec $= -47^\circ22'41.6''$ (J2000), with a total nominal exposure time of 58.9 ks. Data reduction was carried out using the \textit{XMM--Newton} Extended Source Analysis Software (ESAS) within the Science Analysis Software (SAS; version 21.0)\citep{Jansen2001,Struder2001,Turner2001}.\\
Calibration index and observation data files were generated using the tasks \texttt{cifbuild} and \texttt{odfingest}, respectively. EPIC PN and MOS event lists were produced using the \texttt{epchain} and \texttt{emchain} tasks. MOS1 CCD~\#6 was unavailable due to its permanent loss following a micrometeoroid impact on 9 March 2005, and MOS2 CCD~\#4 was excluded because of its unreliable response below 1~keV. Periods affected by soft-proton flares were filtered using \texttt{mos-filter} and \texttt{pn-filter}, and point sources were detected and excised using \texttt{cheese} \citep{Snowden2011}.\\
Cleaned event lists were processed with \texttt{pn-spectra} (\texttt{mos-spectra}) and \texttt{pn-back} (\texttt{mos-back}) to extract source spectra, model the instrumental background, and generate EPIC images in the 0.4--2.0 keV band. The spectra were grouped using \texttt{grppha} from the \texttt{HEASoft FTOOLS} package \citep{Blackburn1995}. 
To model the diffuse X-ray background, which below 2 keV is dominated by soft Galactic emission, we retrieved the appropriate X-ray background spectra, ROSAT All-Sky Survey spectra, and their corresponding response files from the \texttt{HEASARC} X-ray background tool \citep{Snowden1997}. Spectral fitting was performed using \texttt{XSPEC} (version 12.14) \citep{Arnaud1996}, and the residual soft-proton component was modelled using the \texttt{ESAS} \texttt{proton} task.
Finally, an exposure-corrected, background-subtracted image in the 0.4--2.0~keV band was created using the \texttt{ESAS} tasks \texttt{comb} and \texttt{adapt}.

\section{Results \& Discussion}

\subsection{\textbf{Continuum maps morphology}}

\begin{figure}[H]
    \centering
    \includegraphics[width=1\linewidth]{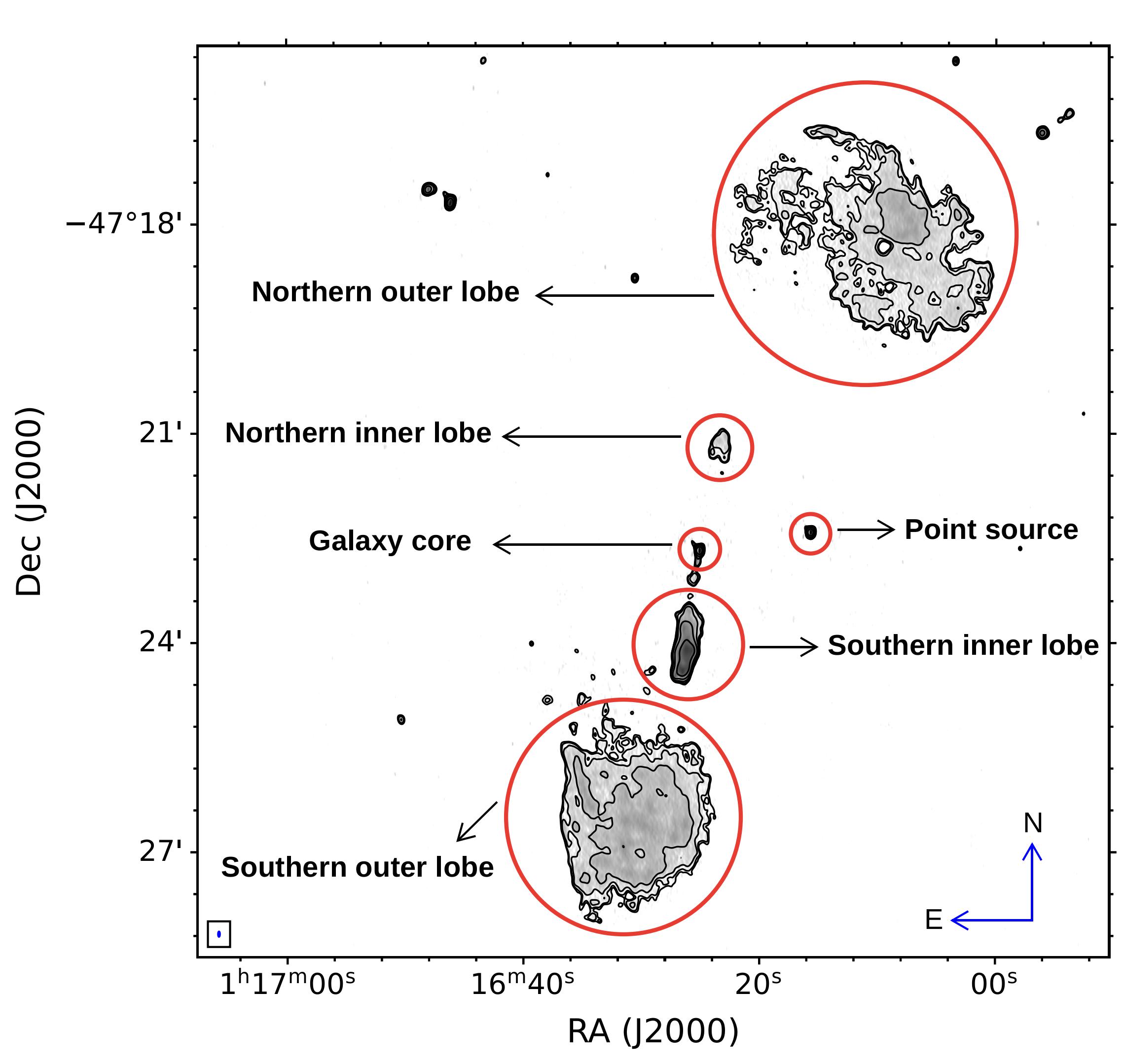}
    \caption{Various regions marked in red circles over the band~5 continuum map. The size of the synthesized beam is indicated by a blue ellipse at the bottom left corner. }
    \label{fig:continuum_map_with_marked_region}
\end{figure}

As discussed earlier, the target source is a double–double radio galaxy (DDRG), characterized by two pairs of radio lobes: an inner pair and an outer pair. Throughout this paper, we adopt the following nomenclature for these components: northern inner lobe, northern outer lobe, southern inner lobe, and southern outer lobe, corresponding to the inner and outer lobes in the northern and southern directions. This can be seen in Fig.~\ref{fig:continuum_map_with_marked_region}.\\

The northern inner lobe of the radio galaxy coincides with the alleged plane of the superdisk. Additionally, we also found a compact source in the plane of the superdisk. If we assume that the superdisk is made up of low-density neutral hydrogen, one would expect a possible H\,{\sc i} absorption, given that the lobes and the compact source are in the background of the superdisk. Motivated by this, we extracted absorption spectra to look for H\,{\sc i} absorption signature (see section~\ref{subsec:HI_absorption_spectra}).

\begin{figure*}[!t]
    \centering

    \begin{minipage}[t]{0.32\textwidth}
        \centering
        \includegraphics[width=\linewidth]{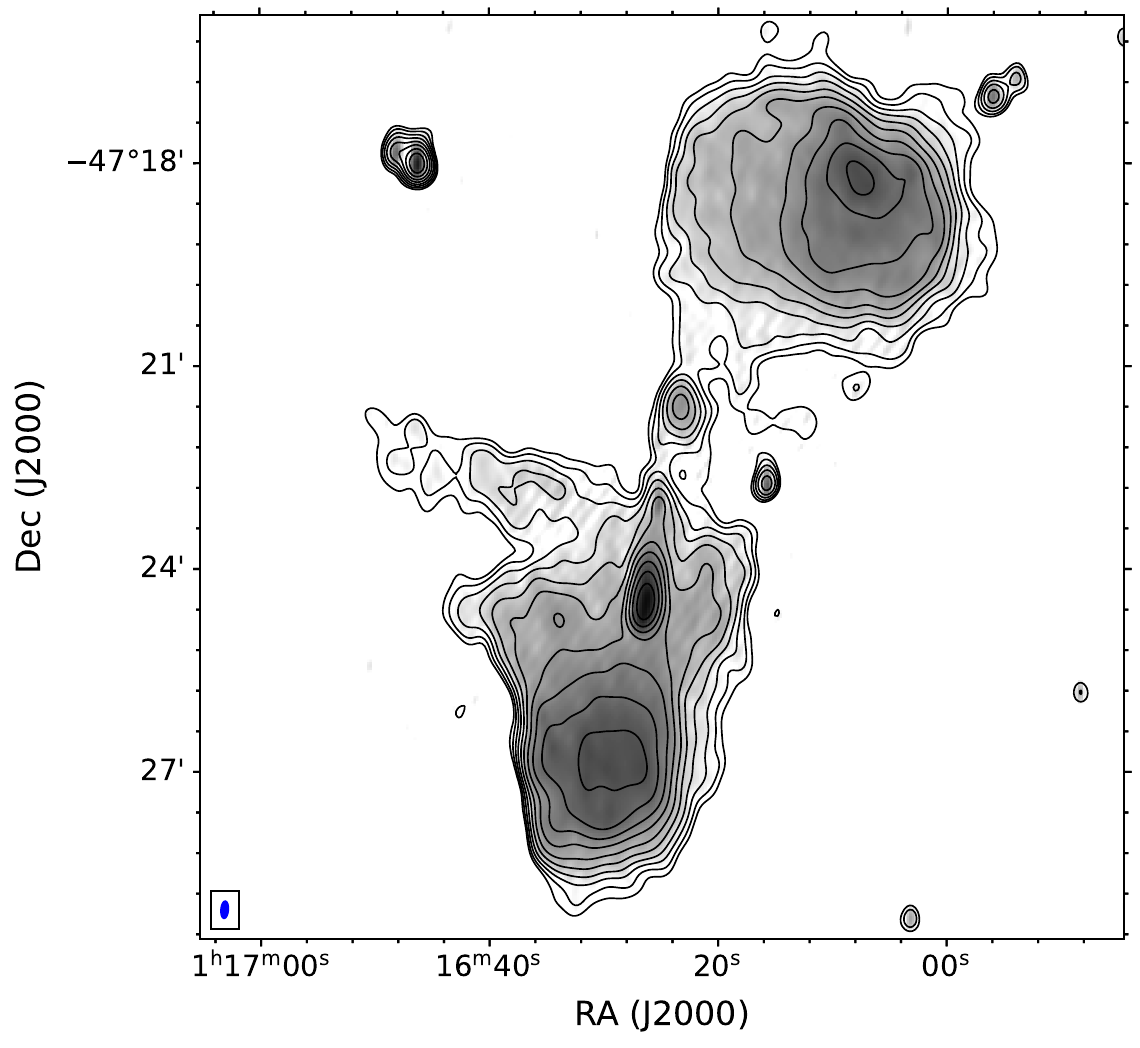}
        \small 
        (a) Band 3 continuum image of J0116-473 at 325 MHz with a beam of $15^{\prime\prime} \times 6.2^{\prime\prime}$ at an rms of $0.2\,mJy$. 
    \end{minipage}
    \hfill
    \begin{minipage}[t]{0.32\textwidth}
        \centering
        \includegraphics[width=\linewidth]{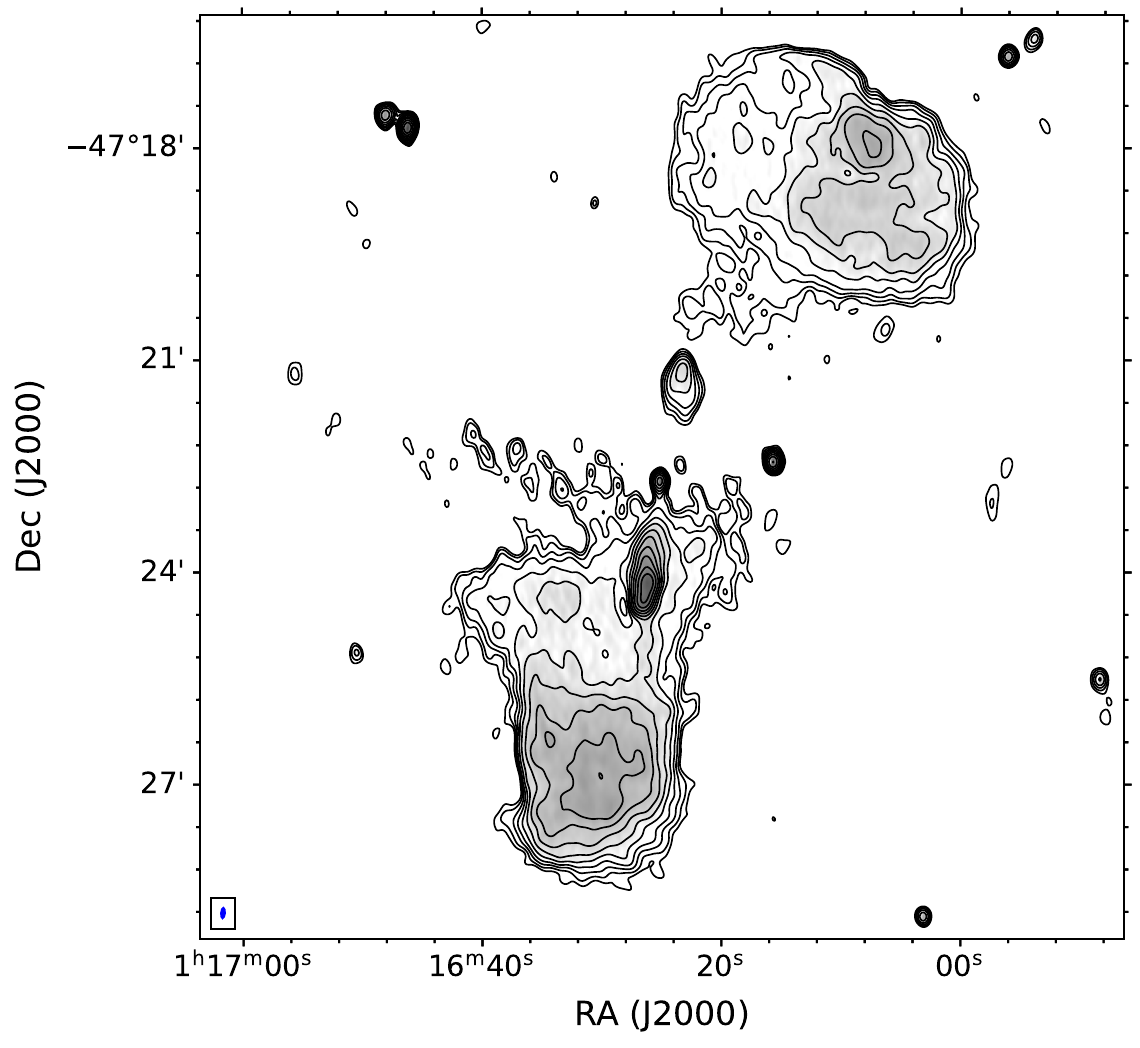}
        \small 
        (b) Band 4 continuum image of J0116-473 at 610 MHz with a beam of $8.6^{\prime\prime} \times 3.1^{\prime\prime}$at an rms of $65\,\mu Jy$.
    \end{minipage}
    \hfill
    \begin{minipage}[t]{0.32\textwidth}
        \centering
        \includegraphics[width=\linewidth]{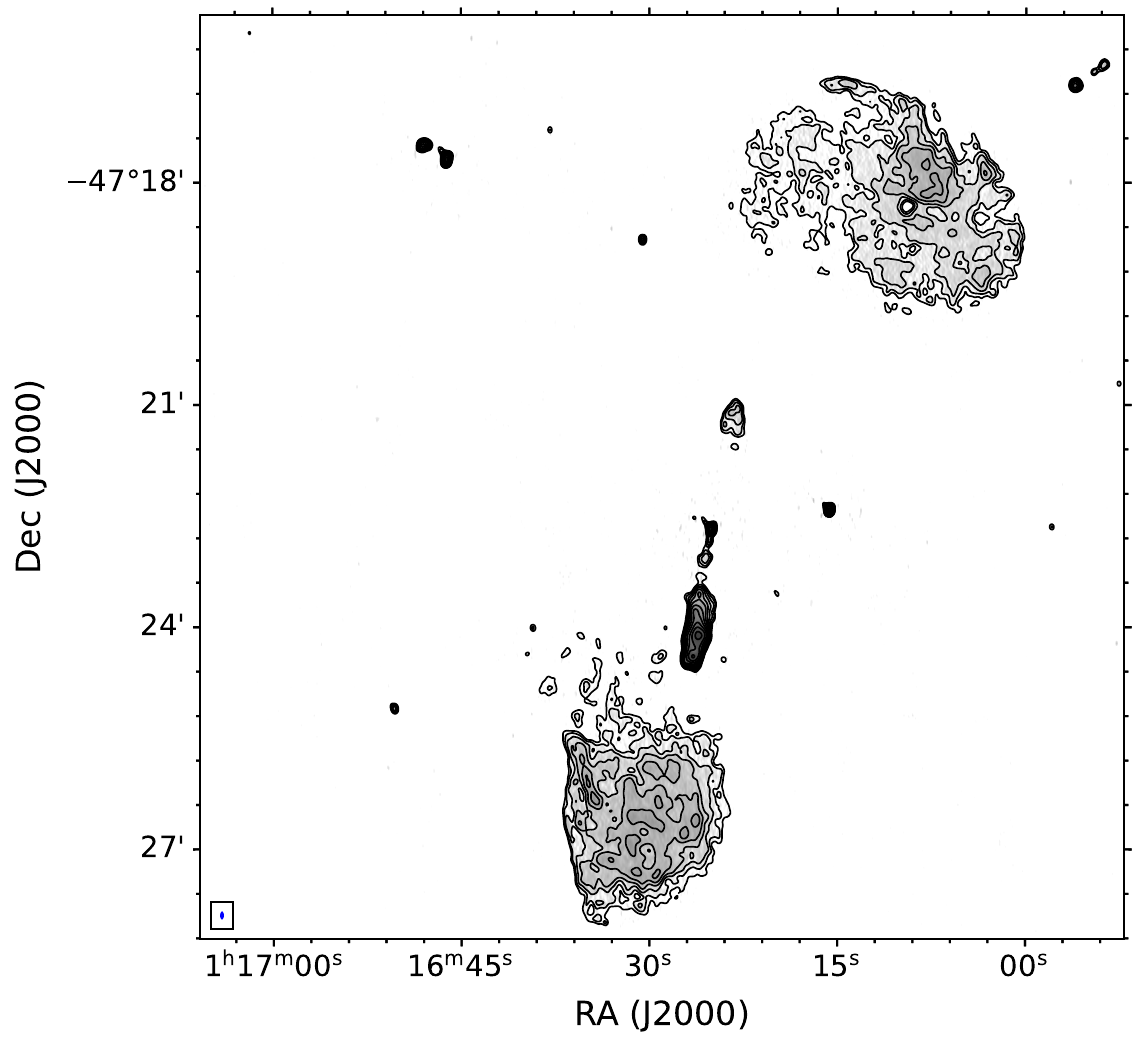}
        \small 
        (c) Band 5 continuum image of J0116-473 at 1240 MHz with a beam of $4.3^{\prime\prime} \times 1.4^{\prime\prime}$ at an rms of $28\,\mu Jy$.
    \end{minipage}

    \caption{Continuum maps of J0116$-$473 at GMRT Bands 3, 4, and 5 with radio contours overlaid. The synthesized beam is shown in blue in the lower-left corner of each panel. The diffuse continuum emission is more prominent in Band 3 than in Bands 4 and 5. For Band 3, contours start at $6\sigma$ and increase by successive factors of $\sqrt{2}$ for 12 levels. For Bands 4 and 5, contours start at $4\sigma$ and $3\sigma$, respectively, and increase by the same factor.}
    \label{fig:continuum_map_morphology}
\end{figure*}

J0116$-$473 is a double--double radio galaxy characterized by two large, diffuse, edge-brightened outer lobes that lack prominent hotspots at their extremities. In addition, a smaller inner double source with brighter lobes shares the same radio core \citep{Saripalli2002}. \citep{Saripalli2002} also reported ``a peculiar elongated structure'' extending nearly a megaparsec in a direction perpendicular to the radio axis of the galaxy and intersecting the axis just south of the core. Their observations were carried out at 1.384 and 2.496~GHz using the ATCA for 12 hours in each of four array configurations, yielding a total integration time of $\approx48$ hours.\\

In this work, we use GMRT observations across three frequency bands to probe the source at lower frequencies than previously reported. In Fig.~\ref{fig:continuum_map_morphology}(c), we can see the band 5 continuum image of the galaxy J0116-473. The image was constructed using $\approx 4.5$ hours of on-source data. We have acheived $28\,\mu Jy$ rms with an observing beam of $4.3^{\prime\prime} \times 1.4^{\prime\prime}$. As shown in the band 5 image, obtained at a central frequency of 1240~MHz, the two large diffuse outer lobes and the smaller inner lobes are clearly visible, with the inner pair nearly coaxial with the outer pair. The Band~5 data provide an opportunity to examine substructures within the radio lobes; however, we do not detect the elongated structure reported in earlier studies at this frequency. \\

In contrast, the Band~3 image, which was constructed using $\approx 4$ hours of on-source data with an rms noise of 0.2mJy, with an observing beam of $15^{\prime\prime} \times 6.2^{\prime\prime}$, reveals a different morphology. An elongated structure is clearly visible near the end of the southern outer lobe (see Fig.~\ref{fig:continuum_map_morphology}(a)). Both the inner and outer pairs of lobes are distinctly detected. As reported by \citep{Saripalli2002}, the outer lobes lack compact hotspots; however, diffuse warm spots are visible near their extremities. 

The southern inner and outer lobes are nearly coaxial, whereas in the northern pair, the outer lobe is slightly misaligned with respect to the inner lobe. A clear intensity contrast is also observed between the southern inner and outer lobes, with the inner lobe appearing significantly brighter. This brightness difference likely reflects the younger age of the inner lobe, consistent with renewed jet activity in the galaxy, as suggested by \citep{Saripalli2002, Konar2013}.\\

The Band~4 image, which was also constructed using $\approx 4$ hours of on-source data, exhibits characteristics intermediate between those seen in Bands~3 and~5. For Band 4 data we acheived an rms noise of $65\,\mu Jy$ with an observing beam of $8.6^{\prime\prime} \times 3.1^{\prime\prime}$. As shown in Fig.~\ref{fig:continuum_map_morphology}(b), this band captures both the diffuse emission and the smaller-scale substructures within the source. The northern inner lobe appears smaller and fainter than its southern counterpart at all three frequencies. This asymmetry is likely due to projection effects, with the southern lobe oriented toward the observer and the northern lobe away from them.\\

As discussed earlier, an elongated structure is visible in the Band~3 image. This provides strong evidence for the presence of a superdisk oriented perpendicular to the radio axis of the galaxy, potentially blocking infalling material from the southern direction towards the core. Another noteworthy feature is the orientation of the superdisk plane. As discussed in Section~\ref{sec:sample_selection}, the northern inner lobe, along with a compact point source, lies within the proposed plane of the superdisk, making H\,{\sc i} absorption studies feasible.


\subsection{\textbf{X-ray morphology of the galaxy J0116-473}}

\begin{figure}[H]
    \centering
    \includegraphics[width=1\linewidth]{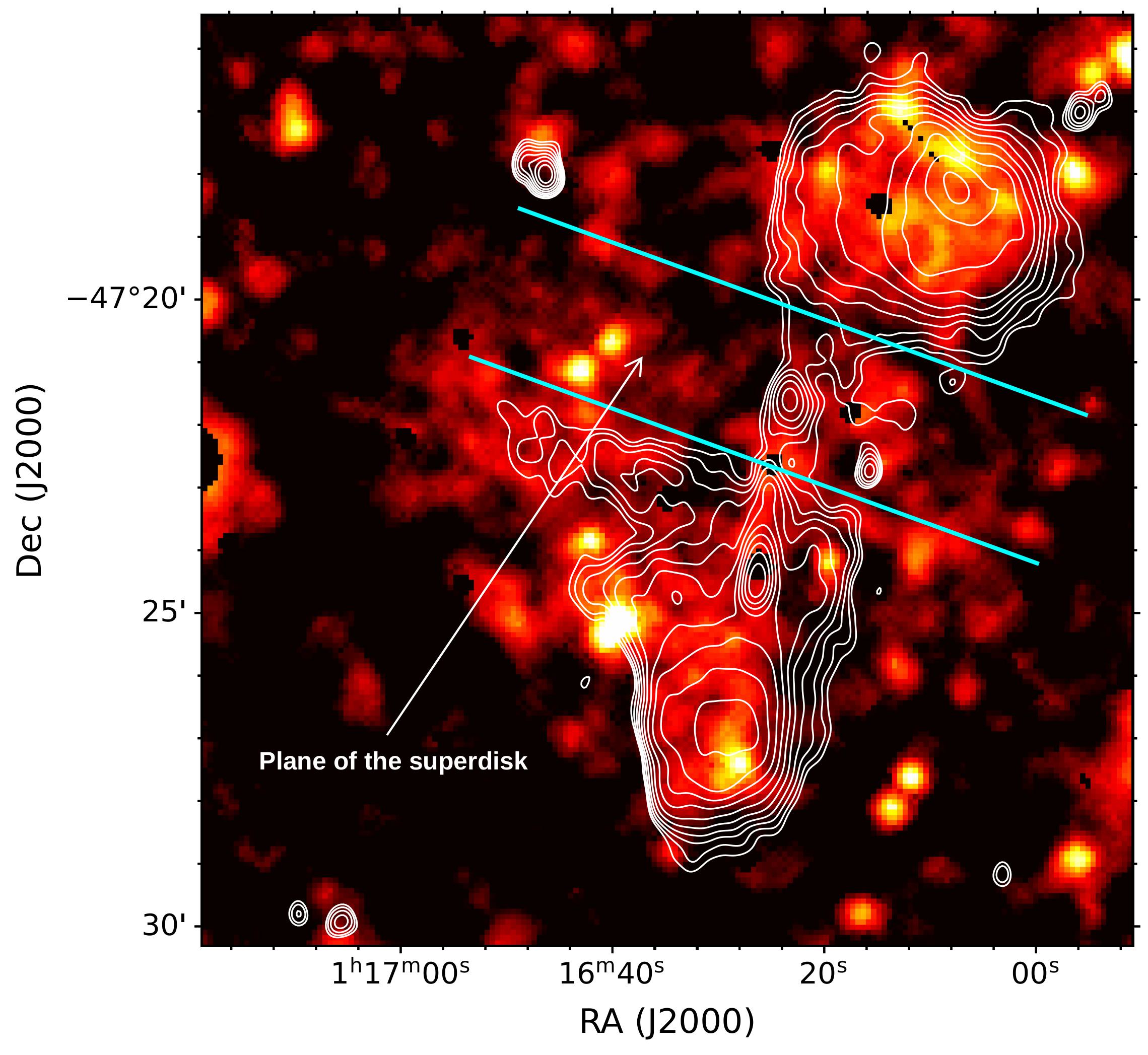}
    \caption{X-ray image of J0116$-$473 derived from XMM–Newton observations, with GMRT Band 3 radio contours overlaid in white. The two parallel cyan lines indicate the plane of the superdisk. The dark region at the position of the host galaxy corresponds to the excised central point source. Radio contours start at $6\sigma$ and increase by successive factors of $\sqrt{2}$ for 12 levels.}
    \label{fig:X-ray_image_with_radio_contour}
\end{figure}

Radio observations reveal the jets and lobes of radio galaxies \citep{Scheuer1974}, but cannot directly probe the hot gaseous atmosphere in which these systems reside. X-ray observations, however, provide important information about the surrounding intragroup or intracluster medium \citep{McNamara2007, Fabian2012}. Using X-ray data, \citep{Duffy2016_3C386} reported the presence of a ``gas belt'' lying between and approximately orthogonal to the radio lobes of the galaxy 3C386. This hot gas could be part of the superdisk and provide a plausible solution to its composition. They measured the temperature of the gas belt to be $\sim 0.94$~keV, which is lower than that of the surrounding extended group atmosphere which has a value of $1.72^{+4.59}_{-0.94}\,\mathrm{keV}$. Furthermore, they identified a temperature gradient within the belt itself, with gas closer to the ambient medium hotter and gas closer to the host galaxy cooler. This temperature structure was interpreted as evidence that the gas represents a relic cold core (for example, see \citep{Sanders2007}).\\
In Fig.~\ref{fig:X-ray_image_with_radio_contour} we show the \textit{XMM--Newton} X-ray image of the galaxy J0116-473, along with the band-3 GMRT contours shown in white. If in J0116-473, the superdisk has a thermal origin, as shown by \citep{Duffy2016_3C386}, we would expect an X-ray emission from that region. However, we do not detect any clear gas-belt-like structure from the region marked by two cyan parallel lines, which represent the plane of the superdisk. Instead, we observe diffuse X-ray emission associated with the outer northern and southern radio lobes. Instead, we observe diffuse X-ray emission associated with the outer northern and southern radio lobes.\\
A possible explanation for this was proposed by \citep{gill2021extended}, who found that the non-thermal X-ray emission associated with the radio lobes dominates over the thermal emission from the surrounding environment. They further argued that this non-thermal emission is primarily produced through inverse Compton scattering of cosmic microwave background (CMB) photons by relativistic electrons in the lobes.

\subsection{\textbf{Characterization of the compact source in the superdisk}}
\label{point_source_classification}

\begin{figure}[H]
    \centering
    \includegraphics[width=1\linewidth]{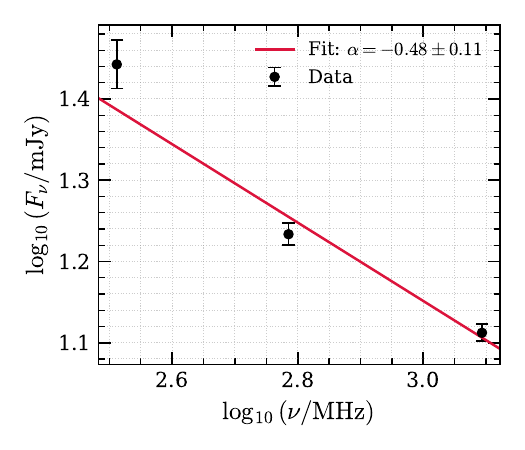}
    \caption{Log-Log plot of frequency vs flux density. We used data from three different frequency bands and performed a linear fit. The unit of flux is mJy, and frequency was measured in MHz. The data points with error bars are shown in black.}
    \label{fig: spectral index of compact object}
\end{figure}
As shown in Fig.~\ref{fig:continuum_map_morphology} and marked in Fig.~\ref{fig:continuum_map_with_marked_region}, a compact, bright point source is observed close to the target galaxy J0116$-$473 and is clearly detected in all three frequency bands. This source lies within the proposed plane of the superdisk associated with the target galaxy. If the source is assumed to lie behind the galaxy, any absorption detected in its spectrum could provide insight into the composition of the superdisk. We therefore search for absorption features in the radio continuum spectrum (see Section~\ref{subsec:HI_absorption_spectra} for details).

However, the relative location of this source with respect to the galaxy is uncertain, as it is not known whether the object lies in the foreground or background. A redshift measurement would resolve this ambiguity; therefore, we conducted a detailed literature search but found no reported redshift for this object. We subsequently searched for the source in several survey catalogs. The object is detected only in radio and infrared bands. The source was detected only marginally in the Wide-Field Infrared Survey Explorer \texttt{WISE} observation \citep{wright2010wide}. Due to this lack of observation, the redshift of the source could not be determined.

To understand the nature of the source, we further performed a spectral index analysis of the point source using the multi-frequency radio data to constrain its nature. The integrated flux density of the source was measured using the \texttt{imfit} task in \texttt{CASA}. These flux densities were then plotted as a function of observing frequency on a log--log scale in Python. A linear fit to $\log(F)$ versus $\log(\nu)$ yields a spectral index of $\alpha = -0.48 \pm 0.11$, indicating a relatively flat spectrum (see Fig.~\ref{fig: spectral index of compact object}).

A flat spectral index in radio galaxies typically indicates optically thin synchrotron emission arising from compact regions such as the AGN core or the base of relativistic jets, where continuous particle injection and, in some cases, Doppler boosting play an important role \citep{Blandford1979,Rybicki1979,Begelman1984}.\\
The faint infrared detection of the point source, together with its non-detection in X-ray, optical, and UV bands, is consistent with obscuration by a dusty torus, which can absorb high-energy radiation and re-emit it at infrared wavelengths \citep{Antonucci1993, Netzer2015, Elitzur2012}.\\
Another similar object, 1413+349, with a radio spectral index of $\alpha=-0.56$ and non-detection in other optical bands except IR, was classified as an optically quiet quasar by \citep{Stickel1996}.

\subsection{\textbf{Spectral analysis of the target source}}

To further understand the nature of the source, we calculated the mean spectral index separately for the core, inner lobe, and outer lobes. The total integrated flux density of each region, along with the error, was measured in each observing band using the \texttt{imfit} task in \texttt{CASA}. The spectral index, $\alpha$, defined by $S_\nu \propto \nu^\alpha$, was determined by performing a linear fit to the $\log S_\nu$--$\log \nu$ relation. The resultant spectral indices are mentioned in table \ref{tab:mean_spectral_index_regions}. These values for the spectral indices are consistent with the values provided by \citep{Saripalli2002}. \\

\begin{table}[h]
\centering
\begin{tabular}{clc}
\hline
Region No. & Region Description & $\langle \alpha \rangle$ \\
\hline
1 & Core              & $-0.27 \pm 0.36$ \\
2 & North lobe (inner) & $-1.56 \pm 0.17$ \\
3 & North lobe (outer) & $-1.50 \pm 0.5$ \\
4 & South lobe (inner) & $-0.92 \pm 0.10$ \\
5 & South lobe (outer) & $-1.41 \pm 0.02$ \\
6 & Compact source     & $-0.48 \pm 0.11$ \\
\hline
\end{tabular}
\caption{Mean spectral index values measured in different regions of J0116--473.}
\label{tab:mean_spectral_index_regions}
\end{table}

The core of the galaxy exhibits a relatively flat spectral index of $\alpha = -0.27 \pm 0.36$, consistent with synchrotron emission from an active galactic nucleus. In contrast, the radio lobes show significantly steeper spectra, indicating an aged population of relativistic electrons that has experienced substantial radiative energy losses.

The northern lobe displays very steep spectral indices in both its inner and outer regions ($\alpha \approx -1.5$), suggesting that it is dominated by old synchrotron plasma. The southern lobe shows a flatter spectrum in its inner region ($\alpha = -0.92 \pm 0.10$) and a steeper spectrum in its outer region ($\alpha = -1.41 \pm 0.02$). This systematic steepening away from the core is consistent with spectral ageing of the electron population due to synchrotron and inverse-Compton losses.

The steeper spectral index observed in the northern inner lobe may be related to its location within the proposed plane of the superdisk. If the superdisk provides a denser confining environment, enhanced radiative and adiabatic losses could lead to more rapid ageing of the relativistic electron population. Although this interpretation is consistent with the observed spectral index asymmetry, the current data do not allow a unique determination of its origin.\\

Overall, the spectral index distribution supports a scenario in which the nucleus remains active while the extended radio lobes contain older plasma. The steep spectra of the outer lobes and the observed spectral steepening with increasing distance from the core provide strong evidence for radiative ageing of the relativistic electrons.

\subsection{\textbf{H\,{\sc i} absorption spectra}}
\label{subsec:HI_absorption_spectra}

\begin{figure*}[h!]
    \centering
    
    \begin{tabular}{cc}
        {\includegraphics[width=0.47\linewidth]{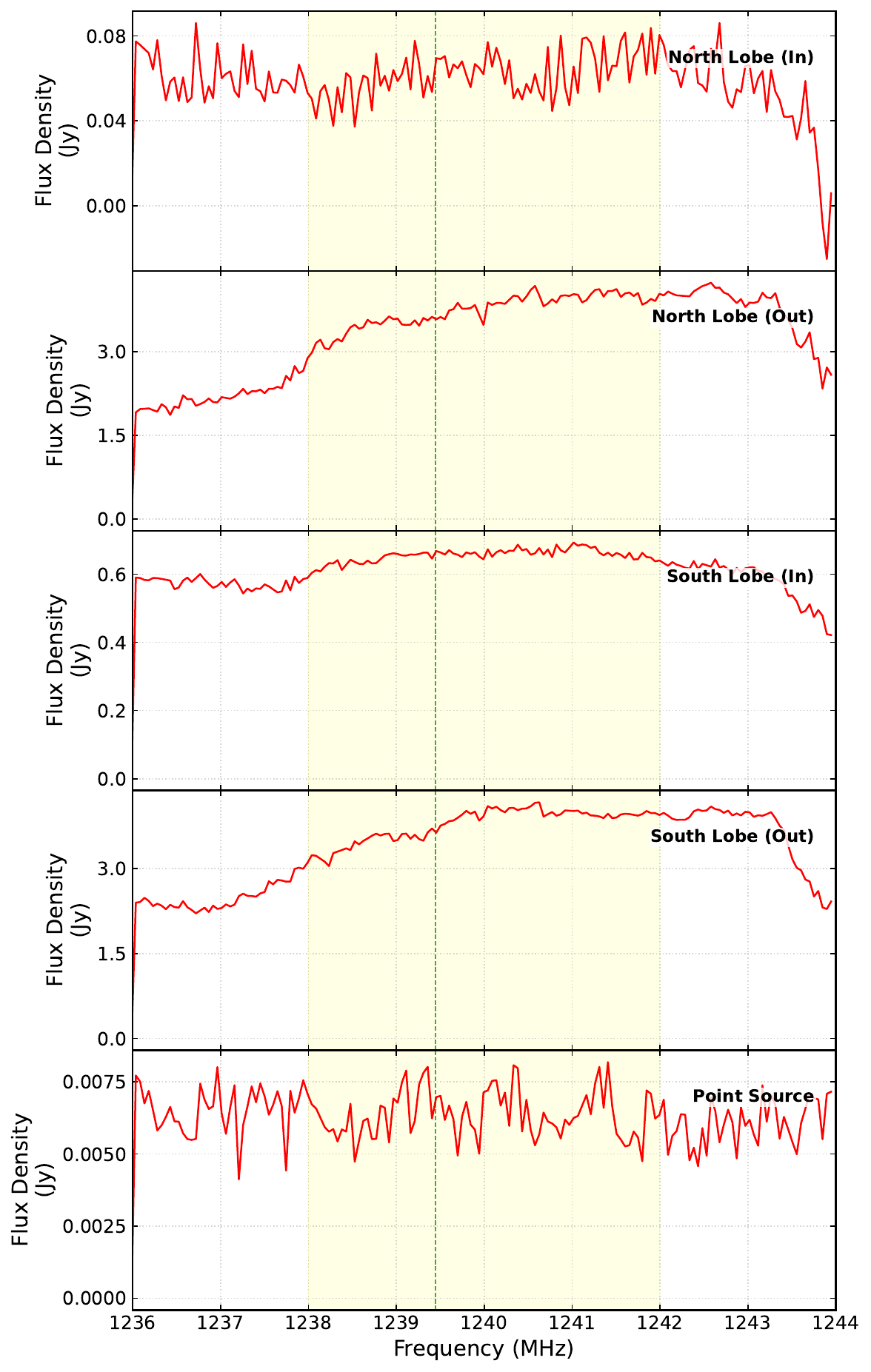}} &
        {\includegraphics[width=0.47\linewidth]{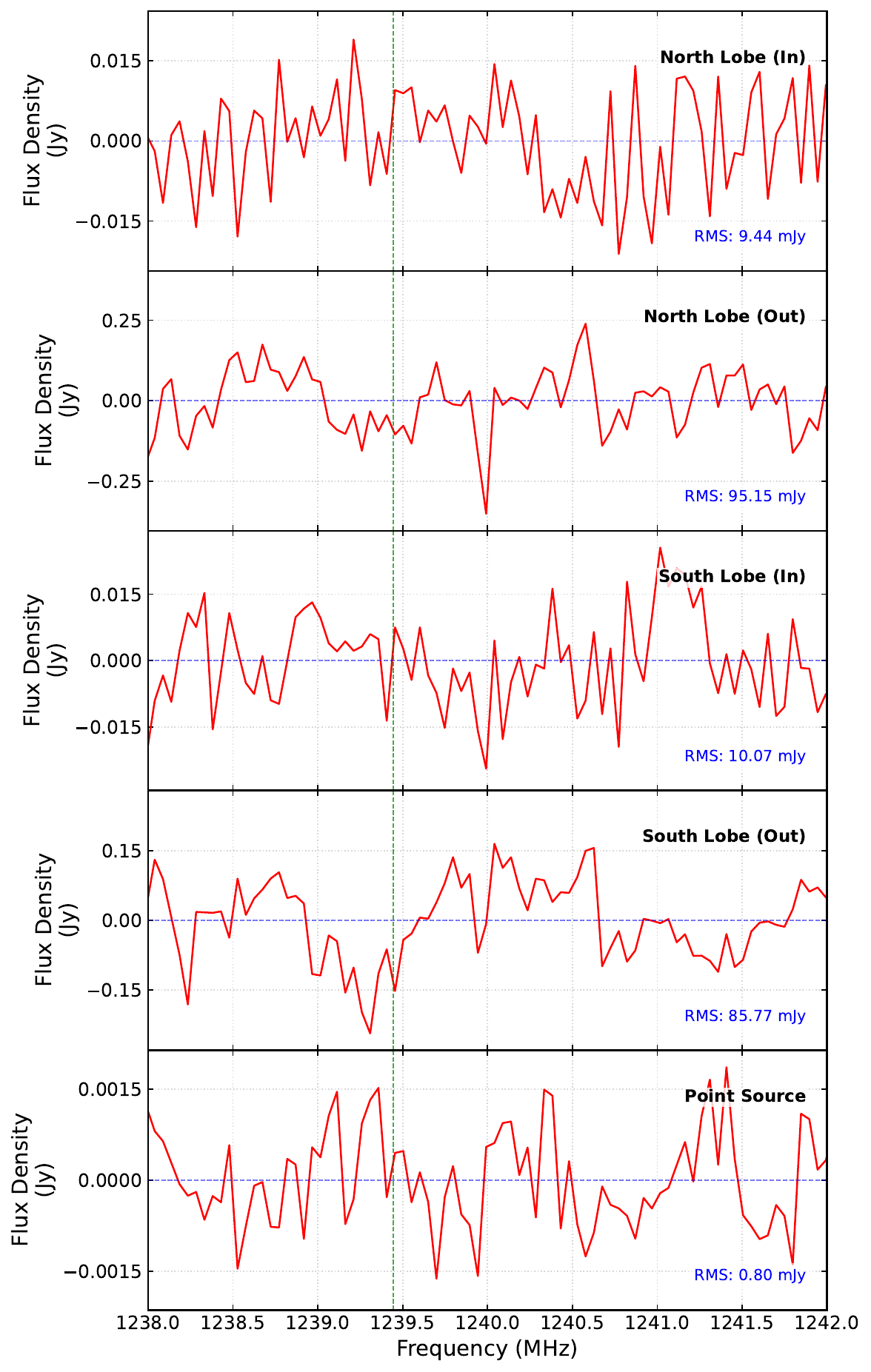}} \\
    \end{tabular}
    \caption{Left column: Flux density vs Frequency plot for the different lobes of the galaxy J0116-473. The dashed green line represents the expected absorption frequency (i.e., 1239.446 MHz); Right column: Flattened Spectrum between the frequency range of 1238 to 1242 MHz, along with the region-specific RMS value. This is the cutout frequency range from the left column highlighted in yellow. The green line represents the expected absorption frequency.}
    \label{tab:spectra_bicolumn}
\end{figure*}

To examine the nature of the superdisk, we constructed the H\,{\sc i} absorption spectrum against the lobes and the compact radio source. If the superdisk contains H\,{\sc i} gas and the compact source resides behind the superdisk, one would expect an absorption feature.\\
Furthermore, we constructed a continuum map of the target galaxy after performing self-calibration. The continuum emission was then subtracted from the visibility data using the \texttt{uvsub} task, and a spectral cube of the target source was subsequently generated using the \texttt{tclean} task in \texttt{CASA} \citep{CASA2022}. Separate \texttt{SAO ds9} \citep{Joye2003} masks were created for different regions of the galaxy. H\,{\sc i} absorption was expected primarily against the northern inner lobe, as this region coincides with the suspected plane of the superdisk. As discussed earlier, for a galaxy at redshift $z = 0.146$, the redshifted H\,{\sc i} 21-cm line is expected at $\approx1239.446$ MHz. Accordingly, the spectral cube was generated by selecting channels around this frequency range.

We then extracted spectra by plotting flux density as a function of frequency in the range 1236–1244 MHz (which corresponds to a velocity width of $\approx1688$ Km/s) for all four lobes of the galaxy (see Fig.~\ref{tab:spectra_bicolumn}). No significant absorption features are detected in any of the spectra. This non-detection may indicate a low or no H\,{\sc i} column density in the superdisk. The spectral shape observed across this frequency range is consistent with the bandpass response, and the spectra are dominated by noise.\\

The first panel in Fig.~\ref{tab:spectra_bicolumn} corresponds to the northern inner lobe of J0116$-$473, as shown in Fig.~\ref{fig:continuum_map_with_marked_region}. The integrated flux density for the northern lobe is 0.18 Jy, which is comparatively lower than that of the other lobes. For the northern outer lobe, the total flux density is 12.28 Jy, for the southern inner lobe it is 2.97 Jy, and for the southern outer lobe it is 13.20 Jy.\\
The rms noise was estimated from the extracted spectrum. We calculated the total flux for each region by summing the channel fluxes using a spatial mask. Since the expected absorption feature lies near 1240~MHz, a spectral window of 4~MHz (1238--1242~MHz), corresponding to a velocity width of $\approx845$ km/s, was selected using a spectral mask. The absence of absorption spectra suggests that the velocity of the H\,{\sc i} in the superdisk is assumed to be less than the above-mentioned value.

Radio spectra often exhibit a curved baseline due to the instrument's bandpass response \citep{thompson2017interferometry}. To remove this effect, a second-degree polynomial was fitted to the spectrum and subsequently subtracted. This baseline subtraction removes any bandpass calibration residual effects caused by the instrument's poor bandpass stability.\\


For the northern inner lobe, against which we attempted to detect H\,{\sc i} absorption, the rms noise is estimated to be 9.44~mJy $beam^{-1} channel^{-1}$. This corresponds to an upper limit on the H\,{\sc i} column density of $3.24 \times 10^{18}\,\mathrm{cm^{-2}}$ at $3\sigma$ level for a continuum flux of $180\, mJy$.

A point source is also present in the suspected plane of the superdisk, see Section~\ref{point_source_classification} for more details. We therefore searched for H\,{\sc i} absorption toward this source using the same procedure described above. Using the spectra of this point source, we estimated an rms noise of 0.80~mJy (see Fig.~\ref{tab:spectra_bicolumn}, last panel).
Using this rms value of 0.80~mJy, the corresponding upper limit on the H\,{\sc i} column density is estimated to be $\approx 2.74 \times 10^{17}\,\mathrm{cm^{-2}}$. This is coming out to be much lower than the upper limit of the column density estimated for the northern inner lobe, because the masked region for the lobe is much larger than that for the point source. \\

We note that the redshift of the above-mentioned point source is currently unknown. A thorough literature search did not yield any published redshift measurements of the source. In the absence of such information, we assume that the point source lies either in the background or at approximately the same redshift as the target galaxy. However, it is possible that the source may be situated at a redshift less than our target galaxy. In that case, we will not be able to see any H\,{\sc i} absorption from the superdisk.

\section{Conclusion}

In this paper, we have studied the radio galaxy J0116$-$473 using GMRT Bands 3 (325 MHz), 4 (610 MHz), and 5 (1240 MHz). Our radio continuum images reveal an elongated structure near the termination of one of the radio lobes, suggestive of the presence of a superdisk in the system. The physical nature and composition of the putative superdisk, however, remain unclear. To investigate the presence of neutral hydrogen associated with this structure, we searched for H\,{\sc i} absorption towards a radio lobe and a compact source located close to the expected plane of the superdisk. No H\,{\sc i} absorption was detected along either line of sight. This non-detection may indicate that the neutral hydrogen column density in the superdisk is below the sensitivity limit of the present observations, or that it is absent.\\


We estimated the upper limits on the H\,{\sc i} column density toward both the targeted lobe and the compact point source. For the lobe, the column density limit is $\approx 3.2 \times 10^{18}\,\mathrm{cm^{-2}}$, while for the point source it is $\approx 2.7 \times 10^{17}\,\mathrm{cm^{-2}}$.\\

We also performed a detailed spectral analysis of the target galaxy by calculating the spectral index from multi-band radio data, enabling us to investigate the physical conditions across different regions of the source. The spectral index distribution exhibits systematic steepening from the core towards the outer lobes, consistent with radiative aging of the relativistic electron population. The core shows a relatively flat spectral index ($\alpha = -0.27 \pm 0.36$), indicative of ongoing AGN activity, whereas the lobes exhibit significantly steeper spectra. The northern inner and outer lobes have spectral indices of $\alpha = -1.56 \pm 0.17$ and $-1.50 \pm 0.50$, respectively, while the southern lobe steepens from $\alpha = -0.92 \pm 0.10$ in the inner region to $\alpha = -1.41 \pm 0.02$ in the outer region. \textit{The steeper spectrum of the northern inner lobe may be associated with its interaction with the proposed superdisk, although the current data do not permit a unique determination of the origin of this asymmetry.}\\

In addition, we analyzed a compact point source near the galaxy core that had been detected in previous surveys but remained unclassified. Using multi-frequency radio data, we determined its spectral index to be $\alpha \approx -0.48$, indicating a relatively flat spectrum. Although the nature of this object remains uncertain, it may represent an obscured active galactic nucleus (AGN), where emission from the central engine is partially absorbed by surrounding gas and dust.

Finally, motivated by the results of \citep{Duffy2016_3C386}, we analysed archival X-ray observations of J0116$-$473 to search for thermal emission associated with a possible gas belt between the radio lobes. In the \textit{XMM--Newton} X-ray images, diffuse emission associated with the outer radio lobes is detected; however, we do not find any clear evidence for a gas-belt-like structure near the ends of the lobes.\\

In summary, superdisks represent an important and intriguing area of study, with the potential to address several fundamental problems in astrophysics. Determining the redshift of the point source would be interesting for constraining the nature of the superdisk in J0116-473. Future work should also focus on observations with higher sensitivity to place tighter constraints on the neutral hydrogen content. In addition, a more detailed X-ray analysis could help identify features such as shocks associated with recurrent jet activity. It would also be valuable to investigate how J0116$-$473 interacts with its surroundings and whether these interactions influence the properties of the superdisk.

\section*{Acknowledgements}

The authors thank the National Centre for Radio Astrophysics (NCRA) for providing the GMRT data used in this study. We acknowledge the Indian Institute of Technology Indore for providing the computational facilities used for the data reduction and analysis. This work used observations obtained with XMM-Newton, an ESA science mission with instruments and contributions funded directly by ESA Member States and NASA.

\vspace{-1em}


\bibliographystyle{mnras}
\bibliography{Draft1/references}

\end{document}